\documentclass[journal]{IEEEtran}
\usepackage{multirow}
\usepackage[table,xcdraw]{xcolor}
\usepackage{booktabs}
\usepackage{graphics}
\usepackage{graphicx}
\usepackage{gensymb}
\usepackage{dblfloatfix}
\usepackage{tikz}
\usepackage{amsmath}
\usepackage{mathtools}
\usepackage{caption}
\usepackage[ruled, vlined]{algorithm2e}
\usepackage[pagewise, switch]{lineno}

\usepackage{soul}

\usepackage{amsmath}
\usepackage{amstext}
\usepackage{amssymb}

\usepackage{fancyhdr}
\newcommand\norm[1]{\left\lVert#1\right\rVert}
\ifCLASSINFOpdf

\else

\fi

\begin{document}

\title{A hybrid approach for dynamically training a torque prediction model for devising a human-machine interface control strategy}

\author{Sharmita Dey$^{1,4,*}$,~\IEEEmembership{}
        Takashi Yoshida$^{1}$,~\IEEEmembership{}
        Robert H. Foerster$^{1}$, Michael Ernst$^{2}$, Thomas Schmalz$^{2}$, Rodrigo M. Carnier$^{3}$, ~Arndt F. Schilling$^{1}$~\IEEEmembership{}%
\thanks{$^{1}$ Applied Rehabilitation Technology Lab (ART-Lab), Department of Trauma Surgery, Orthopedics and Plastic Surgery, University Medical Center Goettingen, Goettingen, Germany e-mail: sharmita.dey@med.uni-goettingen.de; arndt.schilling@med.uni-goettingen.de}%

\thanks{$^{2}$ Research Biomechanics, Clinical Research \& Services, Ottobock SE \& Co. KGaA, Goettingen, Germany
        }%

\thanks{$^{3}$ Yokohama National University, Yokohama, Japan}

\thanks{$^{4}$ University of Göttingen (Georg-August-Universität Göttingen), Germany
        }%
        
\thanks{$^{*}$ Any correspondences should be sent to {\tt\small sharmita.dey [at] med.uni-goettingen.de}}%

}

\maketitle

\thispagestyle{fancy}
\chead{This work has been submitted to the IEEE for possible publication. Copyright may be transferred without notice, after which this version may no longer be accessible.}

\begin{abstract}

Human-machine interfaces (HMI) play a pivotal role in the rehabilitation and daily assistance of lower-limb amputees. The brain of such interfaces is a control model that detects the user’s intention using sensor input and generates corresponding output (control commands). With recent advances in technology, AI-based policies have gained attention as control models for HMIs. However, supervised learning techniques require affluent amounts of labeled training data from the user, which is challenging in the context of lower-limb rehabilitation. Moreover, a static pre-trained model does not take the temporal variations in the motion of the amputee (e.g., due to speed, terrain) into account.
In this study, we aimed to address both of these issues by creating an incremental training approach for a torque prediction model using incomplete user-specific training data and biologically inspired temporal patterns of human gait. To reach this goal, we created a hybrid of two distinct approaches: a generic inter-individual and an adapting individual-specific model that exploits the inter-limb synergistic coupling during human gait to learn a function that predicts the torque at the ankle joint continuously based on the kinematic sequences of the hip, knee, and shank. An inter-individual generic base model learns temporal patterns of gait from a set of able-bodied individuals and predicts the gait patterns for a new individual, while the individual-specific adaptation model learns and predicts the temporal patterns of gait specific to a particular individual. The iterative training using the hybrid model was validated on eight able-bodied and five transtibial amputee subjects. It was found that, with the addition of estimators fitted to individual-specific data, the accuracy significantly increased from the baseline inter-individual model and plateaued within two to three iterations.

\end{abstract}

\IEEEpeerreviewmaketitle

\section{Introduction}

Human locomotion is a complex and coordinated activity of different muscles, limbs, and joints. Lower limb amputations or neuro-muscular impairments severely hinder the locomotive functionality in people. To restore the lost locomotive capability, patients typically use prostheses/orthoses. However, the commercially available prostheses are usually passive \cite{au2008powered} and provide only limited assistance. For instance, during human locomotion, ankle joints must generate appropriate torque to support forward propulsion \cite{donelan2002mechanical, neptune2001contributions}. Besides, walking at different speeds require the ankle joint to produce different torque profiles \cite{winter1983energy}. However, the commercially available passive transtibial prostheses do not contain any active elements that would generate the required torque. Consequently, an amputee needs to compensate for the missing ankle torque by increasing her/his energy expenditure (e.g., by performing exaggerated hip/knee movements), which also results in walking speed that is only at fifty percent of that of an able-bodied individual \cite{waters1976energy}
Therefore, the individuals with the passive replacement  of the ankle joint often exhibit
and unnatural gait asymmetry with its related problems
which leads to increased forces acting on the sound joints and increased oxygen consumption
\cite{waters1976energy, seroussi1996mechanical}. 

A powered/active prosthetic ankle addresses these issues by actively driving the prosthetic ankle joint to mimic the dynamics of a biological limb \cite{herr2012bionic, au2009powered, cherelle2013design, hitt2010active}  by regulating either its velocity or torque. To accomplish this task, the prosthetic device requires a control strategy that detects the user's intention using the sensor inputs measuring her/his current kinetics and kinematics and assigns corresponding control commands (i.e., velocity or torque). for the prosthetic ankle joint to produce the intended locomotion.

Discrete and continuous methods can be utilized to learn a relation between the inputs from the sensors and the control commands to be predicted, to accomplish a certain task \cite{zhuang2019ensemble,hahne2018simultaneous}. Traditionally, control of active lower-limb prostheses is achieved using finite state machines, which use sensor inputs to identify the intended locomotion mode and gait phase discretely, and assign a pre-defined state (kinematics, or kinetics) for the prosthetic device based on the identified phase of the gait using look-up tables \cite{varol2007decomposition, sup2009preliminary, sup2010upslope}. However, since each gait phase and/or locomotion mode is controlled separately, switching heuristics for transitioning between different gait phases and/or locomotion modes need to be defined and look-up tables need to be maintained. Thus, this requires extensive tuning of controller parameters which monotonically increases with finer sections of gait phases and the number of locomotion modes. The parameter tuning was automated to some extent by using pattern recognition algorithms like support vector machines \cite{huang2011continuous}, Gaussian mixture models \cite{varol2009multiclass}, dynamic Bayesian networks \cite{young2014intent}, artificial neural networks \cite{au2005emg, huang2008strategy} for the identification of the gait phase and locomotion modes. However, since the output commands are predefined, this method can support only a limited number of predefined gait phases and/or locomotion modes and lacks variability across gait cycles, and user adaptation.

Another prominent approach in this direction uses the phase-based control \cite{8302866, villarreal2020controlling, embry2018modeling}, which uses sensor inputs to estimate the instantaneous gait phase using trigonometric regression and allocates a corresponding control command to the prosthetic device. Although this method achieves a finer control over the prosthesis compared to the discrete approaches, the use of a static model limits adaptability to temporal variability in gait patterns. 

Recently, data-driven continuous methods based on artificial neural networks \cite{horst2019explaining, ardestani2014human, xiong2019intelligent, lim2020prediction,mundt2020estimation}, support vector regression \cite{dey2019support}, and Gaussian process regression \cite{dhir2018locomotion, yun2014statistical} were also employed to estimate gait variables from sensory inputs. Such regression algorithms allow to learn a relation between the inter-limb motions and predict the kinematics or kinetics of a biological joint from the motion data of a residual limb \cite{dey2019support,mundt2020estimation}. This method could potentially be used to generate biologically inspired trajectories for controlling a prosthetic device according to the sensor inputs from the residual body of the prosthesis user continuously, without discretizing the gait phase into predefined sections. For example, \cite{dhir2018locomotion} proposed Gaussian process regression for learning the power required at the ankle as a function of speed and percentage of the gait cycle. In \cite{ardestani2014human}, the authors used wavelet neural networks to predict joint moments from EMG signals and ground reaction forces. A Gaussian process regression was used in \cite{yun2014statistical} to predict the ankle kinematics from anthropometric features and gait percentage. A support vector regression was used in \cite{dey2019support} to predict both ankle angle and torque using hip and knee angles in able-bodied subjects. In \cite{xiong2019intelligent}, the authors used artificial neural networks to predict the moments at the hip, knee, and ankle joint for able-bodied individuals from six input EMG signals. In \cite{lim2020prediction}, the authors used artificial neural networks to predict the angles and moments of the lower extremity joints of able-bodied subjects from their center of mass kinematics. In \cite{mundt2020estimation}, the joint kinematics and kinetics were estimated for able-bodied participants using inertial measurement unit data from their limb segments using artificial neural networks.

The data-driven continuous methods  can greatly reduce the number of parameters to be tuned, support  variability in gait cycles across different walking trials, and to some extent support non-cyclic movements. However, most of the related works \cite{horst2019explaining}--\cite{yun2014statistical} have proposed pre-trained static models that do not take into account inter-individual variability in gait patterns \cite{allard2017urban, wahid2016multiple, stansfield2003normalisation, hof1996scaling, senden2012importance}, as well as adaptation to temporal changes in the input patterns (e.g., due to terrain or speed) that predict the outputs. 
Another challenge of using the data-driven models for active prosthesis control is acquiring locomotion data from prosthesis users for training the algorithm. On one hand, acquiring a sufficient amount of locomotion data for training the algorithm is a time-consuming process. On the other hand, it would not be possible to obtain the desired reference outputs for controlling the prosthesis from the prosthesis user herself due to missing limb, nor would it be possible to replicate strategies from other amputees due to amputation height, other amputation technique, other muscular fitness, etc).

In this study, we aimed to address these limitations by presenting a hybrid strategy for dynamically training a ankle torque prediction model from biologically inspired temporal patterns of normal and amputee' gait.
To achieve this, ensemble-based predictive models were trained to exploit the inter-limb synergistic coupling in human gait for learning a function that continuously predicts the torque at the ankle joint. In contrast to the usual way of using a static prediction model, we propose dynamically updating a generic base model using individual-specific data by exploiting the flexibility of the ensemble-based models to dynamically add estimators fitted to a pre-trained model. Our results show effective convergence of ankle torques of amputee gait trajectories to normal gaits trajectories. 

The rest of the paper is organized as follows: Section \ref{sec:methods} (Methodology) presents the biomechanical data acquisition procedure, feature extraction, our specific optimization of the prediction strategy and the developed torque prediction model. Section \ref{sec:results} (Results) presents the metrics and methods of evaluation of convergence of the algorithm, with two types of trials: accurate regeneration of ankle torques of normal gaits using reduced kinematic information, and generation of ankle torques for active prosthesis based on inter-individual and individual-specific gait data. Section \ref{sec:discussion} (Discussion) compares the convergence of our methodology with the convergence of other machine learning models and presents the next steps of this research, aiming for an experimental validation. Finally, section \ref{sec:conclusion} (Conclusion) finishes the discussions of this paper.

\section{METHODOLOGY} \label{sec:methods}

\subsection{Ethics} \label{ethics}

The ethics committee of the University Medical Center Goettingen, Goettingen, Germany consented to all the ethics and protocols for our experiments (application number: 18/03/26),  The participants provided their written approval prior to the experiments. 

\subsection{Overview} \label{overview}

\begin{figure*}[!hptb]
    \centering
    \includegraphics[width=\textwidth]{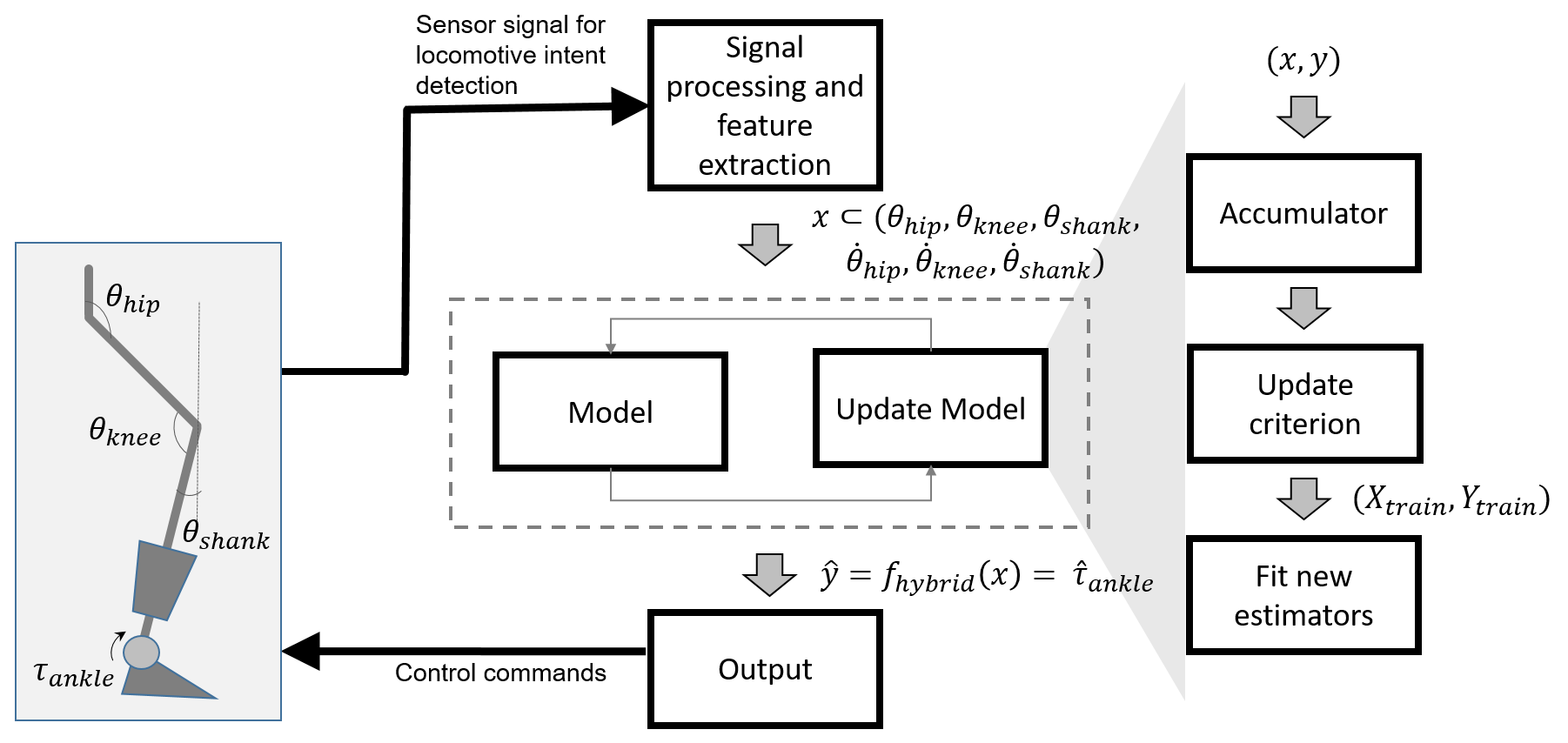}
    \caption{A schematic diagram of the ankle joint torque control framework for an active ankle prosthesis using the proposed hybrid model. The inputs from the prosthesis user obtained using wearable sensors should be processed to acquire the important features to be used as inputs to the model. At each time point, the current model predicts the ankle torque which is used to control the prosthetic ankle. The incoming data from the user can also be accumulated in local storage and used for fitting new estimators to the existing model when an update criterion is satisfied. }
    \label{fig:schematic}
\end{figure*}

Initially, an ensemble-based inter-individual base decision model \cite{breiman2001random, quinlan1986induction} was pre-trained using data from a sample of able-bodied subjects who walked at different ranges of speeds, to capture the generic relationship of kinematic patterns of human gait through reduced degrees-of-freedom of the model (hip, knee, and shank only). Studies have demonstrated the existence of such relationships, naming them movement primitives \cite{Schaal2006}. These proto-solutions have been used before to learn gait models with reduced dimensionality, which are in turn used to predict gait patterns \cite{mukovskiy2017} \cite{carnier2021ml}. While we do not extract movement primitives directly to train our model with them, our pre-trained models possess the same generality of prediction. However, since the usual application of static and generic model has limited success in extrapolating its prediction accurately to a new individual, an adaptation is done to update the model by incrementally adding new estimators, which fit the prediction to individual-specific gait data collected during ambulation. The newly fitted estimators map the kinematics of the amputees’ residual limb (hip, knee, and shank) to normative ankle joint torques, statistically calculated from the walking trials of a group of able-bodied individuals. For further compatibility, the walking speeds of these trials were made similar to the amputees’ particular trials. A schematic diagram of the proposed approach is shown in Fig. \ref{fig:schematic}.

\subsection{Biomechanical data acquisition} \label{biomech-analysis}

We obtained the normative kinematic and kinetic data used in our analyses from a publicly-available dataset \cite{fukuchi2018public}. Data from thirty able-bodied volunteers were selected at five different speeds ($1.04 \pm 0.12$ $m\cdot{s^{-1}}$, $1.22 \pm 0.15$ $m\cdot{s^{-1}}$, $1.4 \pm 0.17$ $m\cdot{s^{-1}}$, $1.58 \pm 0.2$ $m\cdot{s^{-1}}$, $1.77 \pm 0.21$ $m\cdot{s^{-1}}$) as they walked on an instrumented treadmill with a single trial per speed. These five speed levels were selected (out of the eight available speed levels of these subjects, see \cite{fukuchi2018public}), as they subsume the range of speed levels in which our subjects walked during the motion capture experiments (table \ref{tab:info}). Each trial contained one gait cycle, that was characterized by two successive heel strikes. The data from these thirty able-bodied subjects were used to train an inter-individual model as well as to calculate the normative trajectories for amputees (details below). The information about the subjects whose data were used in our analyses is available in table \ref{tab:info}. 

\begin{figure}[!t]
    \centering
    \includegraphics[width = 0.37\textwidth]{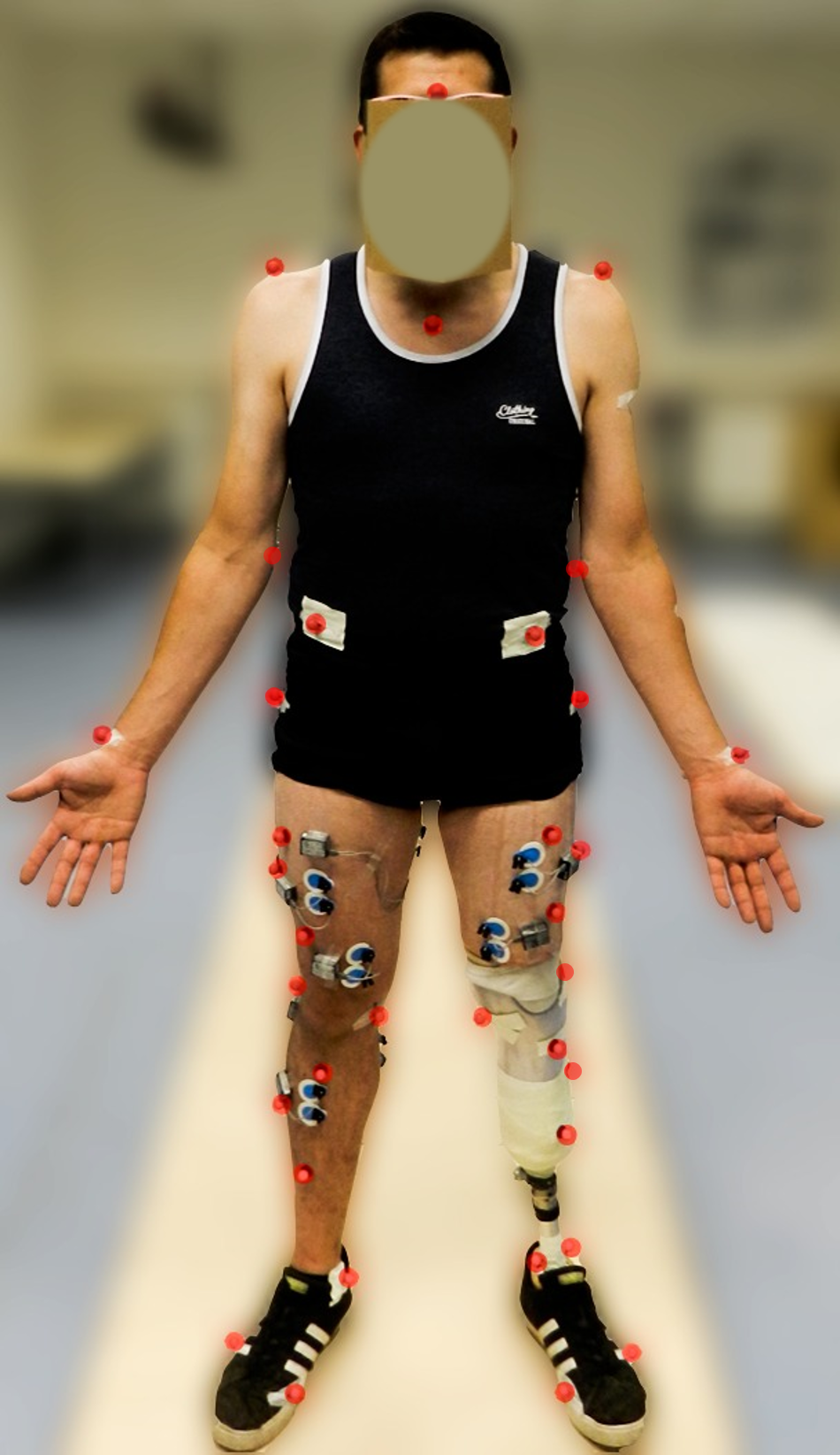}
    \caption{Marker positions (red dots) used during motion capture experiments}
    \label{fig:marker_positions}
\end{figure}

We also recorded 3D motion data from eight able-bodied subjects and five transtibial amputees. Since our experiments were performed on level ground, and not on an instrumented treadmill, it was difficult to vary the walking speed at finer increments. Therefore, the amputees were asked to walk at two different speeds – 1) a self-selected comfortable walking speed, which we called \textit{normal speed}, and 2) a self-selected higher yet comfortable speed, which we call \textit{high speed}.
During all amputee trials, amputees walked using a passive prosthesis based on energy storage and return (see table \ref{tab:info} for information on the prostheses used). We could record normal speed walking trials for all amputees and high speed walking trials for three of the five amputees. Eight able-bodied subjects walked at self-selected comfortable (normal) speeds without wearing any prosthesis. The 3D motion data was recorded using an infrared-based motion capture system (Vicon Motion Systems, Ltd., UK) with retro-reflective markers attached to the bony landmarks at the thorax, lower limbs, and pelvis of the subjects. The position of the markers on a representative amputee subject is shown in Fig. \ref{fig:marker_positions}. Along with the 3D motion data, the ground reaction forces (GRF) were recorded using double force plates (9287A, Kistler Group, Switzerland) at 1000 Hz. 
For each trial, a single gait cycle data, which was characterized by two successive heel strikes on the ipsilateral side (amputated side for amputees or the right leg for able-bodied subjects), was considered for further analyses. 

\begin{table*}
\centering
\caption{Information about data used in our study}
\label{tab:info}
\begin{tabular}{ccccccccccccc} 
\toprule
Subject(s)                                                        & Source                                                                                             & Age                                                   & \begin{tabular}[c]{@{}c@{}}Mass\\(kg) \end{tabular}   & \begin{tabular}[c]{@{}c@{}}Height\\(cm) \end{tabular} & Sex                                                  & Amputation & \multicolumn{2}{c}{\#trials}                                                                      & \multicolumn{2}{c}{\begin{tabular}[c]{@{}c@{}}walking\\speed (m/s) \end{tabular}}                                & \begin{tabular}[c]{@{}c@{}}Prosthesis\\used \end{tabular}                                                                                 & \\ 
\hline
\begin{tabular}[c]{@{}c@{}}30 able-bodied\\subjects \end{tabular} & \begin{tabular}[c]{@{}c@{}}Public\\dataset\\\cite{fukuchi2018public}\end{tabular} & \begin{tabular}[c]{@{}c@{}}39.7 \\±16.8 \end{tabular} & \begin{tabular}[c]{@{}c@{}}67.1\\±~11.5 \end{tabular} & \begin{tabular}[c]{@{}c@{}}168\\±~12 \end{tabular}    & \begin{tabular}[c]{@{}c@{}}16 M/\\14 F \end{tabular} & None       & \multicolumn{2}{c}{\begin{tabular}[c]{@{}c@{}}One trial\\each at\\5 speed \\levels \end{tabular}} & \multicolumn{2}{c}{\begin{tabular}[c]{@{}c@{}}1.04±0.12\\1.22±0.15\\1.4±0.17\\1.58±0.2\\1.77±0.21 \end{tabular}} & None \\ 
\cline{8-11}
                                                                  &                                                                                                    &                                                       &                                                       &                                                       &                                                      &            & normal & fast                                                                                     & normal                                               & fast                                                      &                                                                                                                                           &                                                            \\ 
\cline{8-11}
\begin{tabular}[c]{@{}c@{}}8 able-bodied\\subjects \end{tabular}  & \multirow{6}{*}{\begin{tabular}[c]{@{}c@{}}3D\textasciitilde{}\\motion\\capture \end{tabular}}     & \begin{tabular}[c]{@{}c@{}}26.3\\±~5.0 \end{tabular}  & \begin{tabular}[c]{@{}c@{}}82.6\\± 12.8 \end{tabular} & \begin{tabular}[c]{@{}c@{}}185\\±11 \end{tabular}     & \multirow{6}{*}{M}                                   & None       & 8      & 0                                                                                        & \begin{tabular}[c]{@{}c@{}}1.33\\±0.11\end{tabular}  & --                                                        & None\\
Amputee 1                                                         &                                                                                                    & 62                                                    & 100                                                   & 185                                                   &                                                      & Left       & 8      & 8                                                                                        & \begin{tabular}[c]{@{}c@{}}1.34 \\±0.03\end{tabular} & \begin{tabular}[c]{@{}c@{}}1.56 \\±0.02\end{tabular}      & \begin{tabular}[c]{@{}c@{}}Pro-Flex \\(\textit{Össur}\textasciitilde{}hf., \\Iceland) \end{tabular}                                      \\
Amputee 2                                                         &                                                                                                    & 37                                                    & 91                                                    & 175                                                   &                                                      & Right      & 8      & 8                                                                                        & \begin{tabular}[c]{@{}c@{}}1.22 \\±0.02\end{tabular} & \begin{tabular}[c]{@{}c@{}}1.62 \\±0.04\end{tabular}      & \begin{tabular}[c]{@{}c@{}}1E95 \\Challenger \\prosthesis \\(Otto Bock, \\Germany) \end{tabular}
\\
Amputee 3                                                         &                                                                                                    & 40                                                    & 75                                                    & 178                                                   &                                                      & Left       & 8      & 8                                                                                        & \begin{tabular}[c]{@{}c@{}}1.41\\±0.01\end{tabular}  & \begin{tabular}[c]{@{}c@{}}1.72 \\±0.01\end{tabular}      & \begin{tabular}[c]{@{}c@{}}1C63\\Triton LP\\(Otto Bock,\textasciitilde{}\\Germany) \end{tabular}
\\
Amputee 4                                                         &                                                                                                    &                                                       & 67                                                    & 183                                                   &                                                      & Right      & 8      & 0                                                                                        & \begin{tabular}[c]{@{}c@{}}1.45 \\±0.04\end{tabular} & --                                                        & \begin{tabular}[c]{@{}c@{}}1C68\\Triton \\side flex\\(Otto Bock,\textasciitilde{}\\Germany) \end{tabular}                                 
\\
Amputee 5                                                         &                                                                                                    &                                                       & 80                                                    & 168                                                   &                                                      & Right      & 8      & 0                                                                                        & \begin{tabular}[c]{@{}c@{}}1.29 \\±0.02\end{tabular} & --                                                        & \begin{tabular}[c]{@{}c@{}}1C68\\Triton\textasciitilde{}\\side\textasciitilde{}flex\\(Otto Bock,\textasciitilde{}\\Germany) \end{tabular}
\\
\bottomrule
\end{tabular}
\end{table*}

\subsection{Data processing and extraction of features} \label{sec:section_data_proc}

The marker trajectories and ground reaction forces from the motion capture experiments were  used for  computing the ipsilateral flexion-extension  angles at the hip joint ($\theta_{hip}$) and knee joint ($\theta_{knee}$), forward-backward rotation of the shank segment ($\theta_{shank}$) and the ankle joint dorsi- and plantar-flexion torque (moment) ($\tau_{ankle}$) using open-source biomechanical modeling, simulation, and analysis software (OpenSim \cite{Opensim:Delp:open-source}). Using Opensim, a scaled template subject model, $Z$ with $d$ degrees of freedom, for each subject was created from a generic musculoskeletal model (Gait 2392).  $\theta_{hip}$, $\theta_{knee}$, $\theta_{shank}$ trajectories for each gait cycle were computed by performing  inverse kinematics which solved a least-square convex optimization problem to minimize the deviation between the experimental marker trajectories, $\textbf{m}_i$, and the markers, $\textbf{s}_i(\textbf{q})$, of a musculoskeletal scaled model, $Z$,  for all markers $i$ in $Z$, 

\begin{equation}
    \min_{\textbf{q}}\Sigma_i  \beta_i \norm{\textbf{m}_{i}-\textbf{s}_{i}(\textbf{q})}^2
\end{equation}

where $\boldsymbol{q} \in \mathbb{R}^d$ represents the vector of joint angles, $\beta_i$ represents the respective weights of each marker, $i$. The  $\tau_{ankle}$ trajectories were computed by performing an inverse dynamics which solves the equation of motion

\begin{equation}
    \mathbf{I(q)}\mathbf{\Ddot{q}}+\mathbf{C(q, \dot{q})} + \mathbf{F_g(q)}= \mathbf{\tau}
\end{equation}

where $\mathbf{q}, \mathbf{\dot{q}}, \mathbf{\ddot{q}} \in \mathbb{R}^d$ are the vector of joint angles, angular velocities, and angular accelerations computed from inverse kinematics, $\mathbf{I}(\mathbf{q}) \in \mathbb{R}^{d\times d}$ is the matrix of moments of inertia, $\mathbf{C(\mathbf{q},\dot{\mathbf{q}})} \in \mathbb{R}^d$ is the vector of Coriolis and centrifugal forces, $\mathbf{F_g(q)} \in \mathbb{R}^d$ is the vector of gravitational forces computed from the ground reaction forces, and $\mathbf{\tau} \in \mathbb{R}^d$ is the vector of angular torques. 

\begin{figure*}[!htbp]
    \centering
    \includegraphics[width=\textwidth]{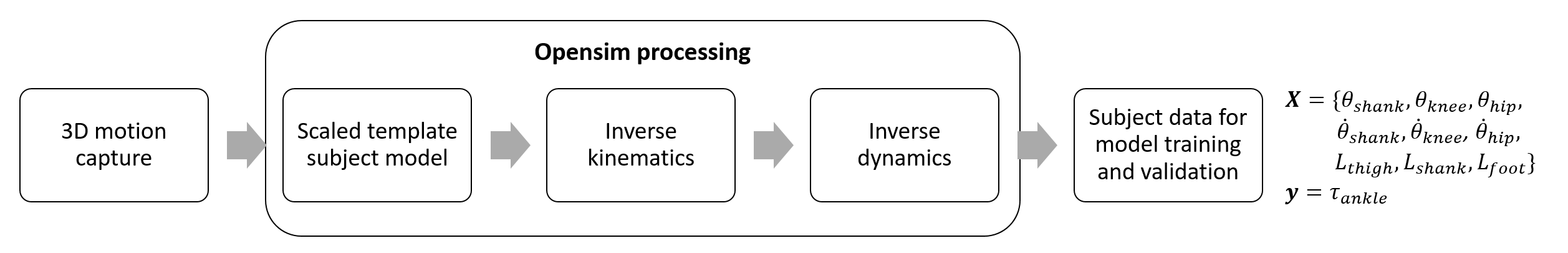}
    \caption{Pipeline used for extracting kinematic and kinetic data for each subject from motion capture data}
    \label{fig:data_processing}
\end{figure*}

The same kinematic and kinetic features ($\theta_{hip}$, $\theta_{knee}$, $\theta_{shank}$, and $\tau_{ankle}$) were used from the public dataset for the thirty able-bodied subjects. The trajectories of these features for each of the gait cycles for able-bodied and amputee were resampled to 200 samples and the ankle moment $\tau_{ankle}$ trajectories were mass-normalized for each subject. Their first derivatives, $\dot\theta_{hip}, \dot\theta_{knee}$, and $\dot\theta_{shank}$ were calculated by taking the numerical difference of $\theta_{hip}$, $\theta_{knee}$, and $\theta_{shank}$ respectively between successive samples. Finally, all these features were low-pass filtered using a Butterworth filter with a cut-off frequency of 6 Hz. Along with the kinematic and kinetic features, we also obtained the anthropometric features like the lengths of thigh $L_{thigh}$, shank $L_{shank}$, and foot $L_{foot}$ segments of the subjects. The data processing pipeline is shown in Fig. \ref{fig:data_processing}. 

The kinematic and kinetic gait features obtained from  the public gait database and our experimental data were acquired analogously by performing marker-based motion capture experiments followed by inverse kinematics and inverse dynamics on the captured motion data using a template subject model. Furthermore, previous studies have reported qualitative and quantitative similarities of the treadmill and overground gaits \cite{riley2007kinematic}. Thus, the public dataset can be considered comparable to our experimental data.

\subsection{Optimization of hyper-parameters and ablation study}

\begin{figure*}
    \centering
    \includegraphics[width=0.8\textwidth]{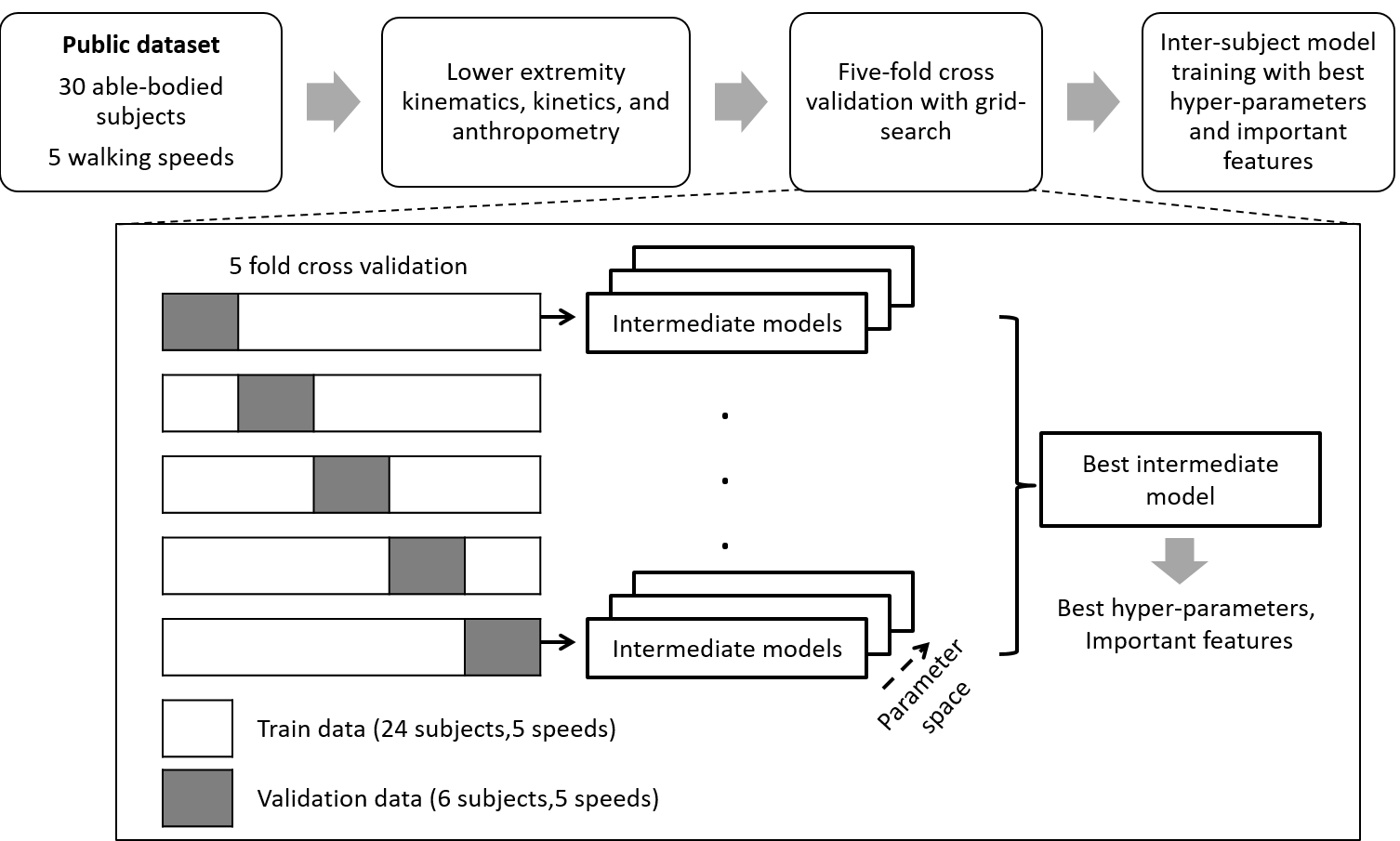}
    \caption{Baseline inter-individual model selection and training. }
    \label{fig:cross_validation}
\end{figure*}

To determine the best hyper-parameters, maximum depth of the estimators, $d_{max}$, and number of estimators, $P$, of the ensemble model, we performed a grid search  on the training set with five-fold cross-validation and a grid of parameter values given as
\begin{equation}
    \label{eq:gridsearch_paremters}
    \begin{aligned}
        d_{max} \in \{4, 6, 10, 20, 50, 100\} \\
        P \in \{10, 20, 50, 100, 500, 1000\}
    \end{aligned}
\end{equation}
For each value in the grid, the ensemble-estimators were trained repeatedly on different subsets (training set) of the original training data and the performance was validated on the remaining subset (validation set). During each iteration, the training set consisted of data from all five speed levels of 24 out of 30 subjects, and the validation set consisted of data from all five speed levels of the remaining six subjects. This was done to train a generalized model that could be used for torque prediction in new subjects at different speeds. The hyper-parameter values which gave the highest mean prediction performance on the validation set across the cross-validation iterations were used as the best hyper-parameters of the prediction model (Fig. \ref{fig:cross_validation}). 

To select the most informative input features, %
an ablation study and feature importance analysis \cite{breiman1984classification} was performed using the data from 30 able-bodied individuals available in the public dataset with all the six kinematic and three anthropometric features (see \ref{sec:section_data_proc}) provided as input. The three most important input features which gave the highest prediction accuracy on the out-of-sample training data were used as input features for further experiments. Selecting only the important features reduces the input dimensionality and thus saves computational cost in real-time applications of prosthesis control. 

\subsection{Ensemble-based torque prediction models}

In this study, the kinematic and anthropometric input features, $\mathbf{x}$ = ($\theta_{hip}$, $\theta_{knee}, \theta_{shank}, \dot\theta_{hip}, \dot\theta_{knee}$, $\dot\theta_{shank}$, $L_{thigh}$, $L_{shank}$, $L_{foot}$) were mapped to the target variable, $y = \tau_{ankle}$ by training ensemble of tree-based decision models. It should be noted that in contrast to some other studies \cite{yun2014statistical}, the time information (instantaneous gait phase percentage) was not explicitly encoded as an input feature in our method, but is implicitly contained in the trajectories of input features.

\subsubsection{Inter-individual model}
An inter-individual base model learns a generalized relation, $f^{II}$, between input features and target variable ($\tau_{ankle}$) using training data from a sample of able-bodied subjects and applies this relation to predict the target variable for a new individual. The relation learned by the inter-individual model can be represented as 

\begin{equation}\label{eqn:inter-individual}
    \min_{f^{II}(H_0)}\biggl(\norm{{\bigcup_{w=1}^W}{\bigcup_{j=1}^J}\mathbf{y}_{w,j}^{AB} - f^{II}\biggl(H_0, {\bigcup_{w=1}^W}{\bigcup_{j=1}^J}\mathbf{X}_{w,j}^{AB}\biggr)}_1\biggr)
\end{equation}

where \(H_0 = (d_0, n_0)\) are the hyperparameters of the model (maximum depth of estimators, \(d_0\), and the number of estimators, \(n_0\)),  \(\mathbf{y}_{w,j}^{AB} \in \mathbb{R}^N\) is the desired \(\tau_{ankle}\) trajectory in a gait cycle (having $N$ data points) of \(j\)-th trial of the \(w\)-th able-bodied subject, \(\mathbf{X}_{w,j}^{AB} \in \mathbb{R}^{N\times{3}}\) are the trajectories of the three most important input features within the gait cycle (having $N$ data points) of the $j$-th trial of the $w$-th able-bodied subject and \(f^{II}\) is an ensemble-learning-based \textit{inter-individual model}. 

\subsubsection{Individual-specific model}
The individual-specific model learns a relation between the input features and the corresponding target variable using data from one or more trials of a specific individual. For example, an individual-specific model, $f_k^{IS}$ with hyperparameters $H_k = (d_k, n_k)$ which maps the input features, $\mathbf{X}_k^{amputee}$, of the subject{'}s $k$-th trial to corresponding output variables, $\mathbf{y}_{k}^{subject}$,  is given by

\begin{equation}
\label{eq:individual-specific}
    \min_{f^{IS}_k(H_k)}\biggl(\norm{\mathbf{y}_{k}^{subject} - f_{k}^{IS}(H_k, \mathbf{X}_k^{subject})}_1\biggr). 
\end{equation}

For an able-bodied subject, the desired target output variables trajectory was measured from the subject her/himself whereas for an amputee, the target desired output trajectory was computed as a median of the temporal gait patterns of target variables from a sample of able-bodied subjects. For an amputee, the normative trajectory $\mathbf{y}^{normative}_k$ for his $k$-th trial was calculated as
\begin{equation}
\label{eq:normative_output}
    \mathbf{y}^{subject}_{k} = \mathbf{y}^{normative}_k =    \textrm{median}(\mathbf{y}^{AB}_{w,j_k}), w \in \{1 \dots W\} ,
\end{equation}
where $\boldsymbol{y}^{AB}_{w,j_k} \in \boldsymbol{R}^N$ is the measured $\tau_{ankle}$ trajectory of $w$-th able-bodied subject's $j_k$-th trial, in which the mean pelvis velocity of the able-bodied subject was within $\pm0.1$ $m\cdot{s^{-1}}$ of that of amputee's $k$-th trial.
In the case of an amputee, the individual-specific model, $f^{IS}_k$, tries to minimize the deviation between the predicted trajectory and the normative trajectory for $k$-th trial and is given by:
\begin{equation}
    \min_{f^{IS}_k(H_k)}\biggl(\norm{\mathbf{y}_{k}^{normative} - f_k^{IS}(H_k, \mathbf{X}_k^{amputee})}_1\biggr)
\end{equation}

\subsubsection{Hybrid model}
The hybrid model can be described as a convex combination of the above two models: a baseline inter-individual model and individual-specific models. The hybrid model is initialized as an inter-individual model, $f^{II}(H_0)$, until new candidate estimators are dynamically combined with the existing model. When an update criterion was satisfied, new estimators fitted to the individual-specific data were added to the ensemble model. An update criterion can be defined in different ways like completion of a gait cycle or trial, or, prediction accuracy falling below an acceptable level.
In this study, the prediction accuracy was used as the update criterion. New estimators were fitted to the individual-specific data at the end of a trial if the mean prediction accuracy for the complete trial was less than a threshold value.

During each update iteration $k \in \{1\dots L-1\}$ (where $L$ is the total number of trials of the subject), the hybrid model is represented as a function:

\begin{equation}
\label{eq:hybrid_model}
    f_k^{hybrid} = 
    \begin{cases}
    f^{II}(H_0) &\text{if k=0}\\
    \alpha_k \cdot f_{k}^{IS}(H_k) + \\\quad(1-\alpha_k)\cdot f_{k-1}^{hybrid} &\text{if}\ R^2(\hat{\mathbf{y}}_{k}, \mathbf{y}_{k}) < \zeta \\
    f_{k-1}^{hybrid} &\text{otherwise}
    \end{cases}
\end{equation}
where 
$\mathbf{y}_k$ and $\hat{\mathbf{y}}_k$ are respectively the actual (read normative for amputees) and predicted $\tau_{ankle}$ trajectory for the $k$-th trial, and $\zeta$ is the threshold value. The threshold, $\zeta$, determines a trade-off between the prediction time complexity and performance of the model. $\alpha_k$ is the ratio of the number of newly fitted estimators in the $k$-th update iteration to the total number of estimators in the model. 

\begin{equation}
    \label{eq:alpha_k}
    \alpha_k = \frac{n_k}{\Sigma_{i = 0}^{k}n_i} 
\end{equation}

where $n_i$ is the number of estimators added to the hybrid model during $i$-th training iteration ($n_0$ indicates the number of estimators in the baseline inter-individual model). Therefore, by adjusting the number of newly fitted estimators, it is possible to adjust the effect of the new training data on the model as well as to vary the inter-individuality and individual specificity of the model. 

\begin{figure*}[!ht]
    \centering
    \includegraphics[width =  \textwidth]{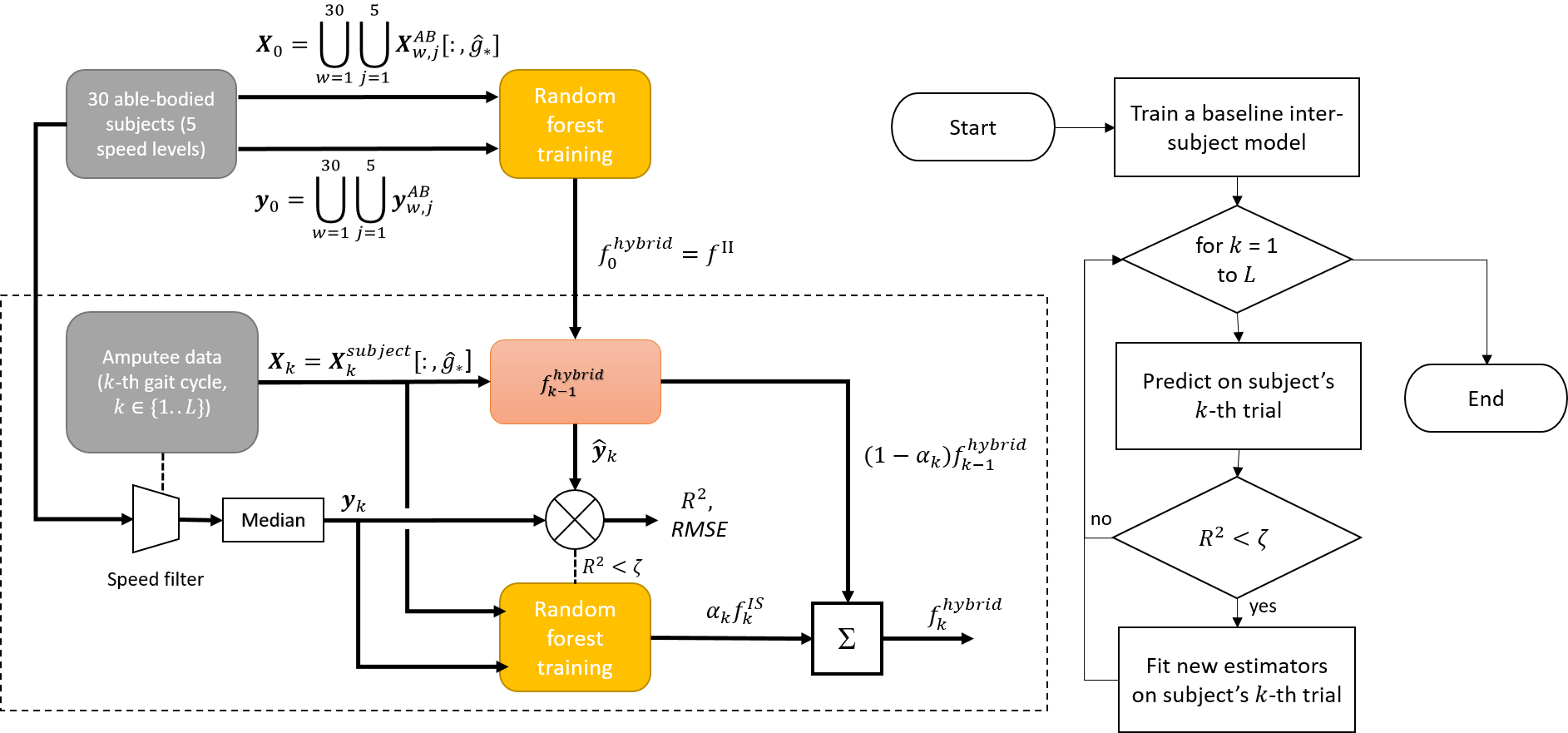}
    \caption{Schematic diagram of the train and validation procedure of the hybrid model for the amputee subjects. The baseline inter-individual model was trained using data from 30 able-bodied individuals who walked at five distinct ranges of speeds. At $k$-th training iteration ($k = 1...L$), data from $k$-th individual-specific gait cycle of walking at self-selected normal speed were used for training new estimators ($L$ is the total number of individual-specific normal speed gait cycles for a subject), and the remaining ($L-k$) normal speed gait cycles and $M$ fast walking gait cycles were used for validation. }
    \label{fig:hybrid_block}
\end{figure*}

The training procedure of a hybrid model used in this study is schematically represented in Fig. \ref{fig:hybrid_block} and can be summarized as follows: 
\begin{enumerate}
   
    \item Find the important features, $\hat{g}_*$, and optimal hyper-parameters, $H_* = (d_*, n_*)$, of the baseline inter-individual model using five-fold cross-validation on data $(\mathbf{X},\mathbf{y}):= \bigcup_{w=1}^W\bigcup_{j=1}^{J}(\mathbf{X}^{AB}_{w, j}, \mathbf{y}^{AB}_{w, j})$ from five speed levels of 30 able-bodied subjects and a grid of hyper-parameter values (see eq. \ref{eq:gridsearch_paremters}).
    \item Set the input data and hyper-parameters of the baseline inter-individual model:
    $H_0 = (d_0, n_0) = H_*$, 
    $\mathbf{X}_0, \mathbf{y}_0 = \mathbf{X}[:, \hat{g}_*], \mathbf{y}$
    \item Fit baseline inter-individual model.
    $f_{II}(H_0)$ using $(\mathbf{X}_0, \mathbf{y}_0)$ as training data and hyper-parameters $H_0$
    \item Initialize the hybrid model, $f^0_{hybrid} := f_{II}(H_0)$
    \item For each normal speed walking trial, $k$, of the subject, 
    \begin{enumerate}
        \item Get the individual-specific training data for the $k$-th trial. 
        $(\mathbf{X}_k, \mathbf{y}_k) := (\mathbf{X}_k^{subject}[:, \hat{g}^*], \mathbf{y}_k^{subject})$. For an amputee $\mathbf{y}_k^{subject} = \mathbf{y}_k^{normative}$
        \item Predict the $\tau_{ankle}$ for the input features of the $k$-th trial of the subject. 
        $\hat{\mathbf{y}}_k^{subject} = f_{k-1}^{hybrid}(\mathbf{X}_k^{subject})$
        \item If the $R^2(\mathbf{y}_k^{subject}, \hat{\mathbf{y}}_k^{subject}) < \zeta$, fit an ensemble of estimators $f^{IS}_k(H_k)$ using the individual-specific data, $(\mathbf{X}_k^{subject}, \mathbf{y}_k^{subject})$, for the $k$-th trial and hyper-parameters $H_k = (d_k, n_k)$. 
        \item Update the hybrid model according to equation \ref{eq:hybrid_model}.\\
        
    \end{enumerate}
\end{enumerate}

Here, we chose a threshold value, $\zeta$, of 0.99. Additionally, we chose $n_k = n_0$ such that the prediction at a new point, $r{’}$ is evenly influenced by both the inter-individual and individual-specific decision when only little individual-specific data is available and swiftly converges to an individual-specific decision as more individual-specific data becomes available. Also, $d_k$ was chosen to be larger than $d_0$ since the deeper trees (estimators) might learn the less variant individual-specific gait patterns more efficiently as opposed to shallow trees which help in generalizing to inter-individual data. 

To ensure that the performance of the hybrid model was not affected by the order of the individual-specific gait data, the gait cycle data was shuffled such that the data at the last index reaches first and all other indices are shifted by one. For each of the shuffled datasets, the incremental validation and training procedure in step 4 was performed. This process was repeated until each of the gait cycle indices reached first at least once. 

Additionally, for each of the new individual-specific normal speed walking gait cycle data added to the training, the model{’}s prediction performance was also evaluated on fast speed walking gait cycles (available only for amputees). Since the model was not trained with any of the fast walking data,  the prediction accuracy on the fast walking data helps to evaluate the model's ability to scale its learning to faster speeds. 

This procedure was followed for all the test subjects (eight able-bodied and five amputees) whose gait data were obtained experimentally in our gait laboratory. Two prediction accuracy metrics were adopted for quantifying the performance of the hybrid model: coefficient of determination ($R^2$) and root-mean-square error (RMSE) between the predicted output trajectory,  $\hat{\mathbf{y}}_{k}^{subject}$, and the desired output trajectory, ($\mathbf{y}_{k}^{subject}$ for able-bodied or $\mathbf{y}_{k}^{norm}$ for an amputee, see eqn. (7)), across a complete trial (one gait cycle in our case). These are given by:

\begin{equation}
    R^2(\mathbf{y}, \mathbf{\hat{y}}) = \frac{\Sigma_{i=1}^N(y_i - \hat{y}_i)^2}{\Sigma_{i=1}^N(y_i - \overline{\mathbf{y}})^2} 
\end{equation}
where $\overline{\mathbf{y}} = \frac{\Sigma_{i=1}^N y_i}{N}$
and 

\begin{equation}
    \text{RMSE}(\mathbf{y}, \mathbf{\hat{y}}) = \sqrt{\frac{\Sigma_{i=1}^N(y_i - \hat{y}_i)^2}{N}}
\end{equation}

The mean $R^2$ and RMSE across all the predicted trials are reported. A Wilcoxon-signed-rank test was performed to compare the performance of the hybrid model during each successive iteration of the individual-specific training. The performance of the hybrid model after the seventh individual specific iteration , $f_7^{hybrid}$ was compared with that of the baseline inter-individual model, $f_0^{hybrid}$, and a pure individual-specific model, $f_7^{subject}$ trained using seven individual specific trials and validated on the remaining trial using leave-one-out cross validation. 
Finally, we also report the time complexity required for training and prediction using the hybrid model for various threshold values for updating the model. The results of this study were obtained using a laptop PC (Intel i7 processor with 16GB RAM and 2.59GHz clock frequency).

\section{Results} \label{sec:results}

A feature importance analysis on data of 30 able-bodied subjects walking at five different speeds revealed that four kinematic features, $\theta_{hip}$, $\dot{\theta}_{knee}$, $\dot{\theta}_{shank}$, and $\dot{\theta}_{hip}$, accounted for 98\% of the total importance among all the six kinematic and three anthropometric features in estimating the ankle torque. Using only these four important features gave comparable accuracy as using all the kinematic and anthropometric features. 
Therefore, we decided to use only the four kinematic features ($\theta_{hip}$, $\dot{\theta}_{knee}$, $\dot{\theta}_{shank}$, and $\dot{\theta}_{hip}$) in our further analysis.  

Using the five-fold cross-validation of the ensemble prediction models on the public dataset, the best hyper-parameters were obtained to be a maximum depth, $d_*$ of 6, and the number of estimators, $n_*$ = 100. These were used as hyper-parameters, $H_0 = (d_0, n_0)$ for training the baseline inter-individual model. For each subsequent individual-specific training iteration $k$, the same number of estimators, $n_k = n_0$, were added to the model, but having a maximum depth, $d_{k}$ of 10. 

\begin{figure*}
    \centering
    \includegraphics[width = 0.98\textwidth]{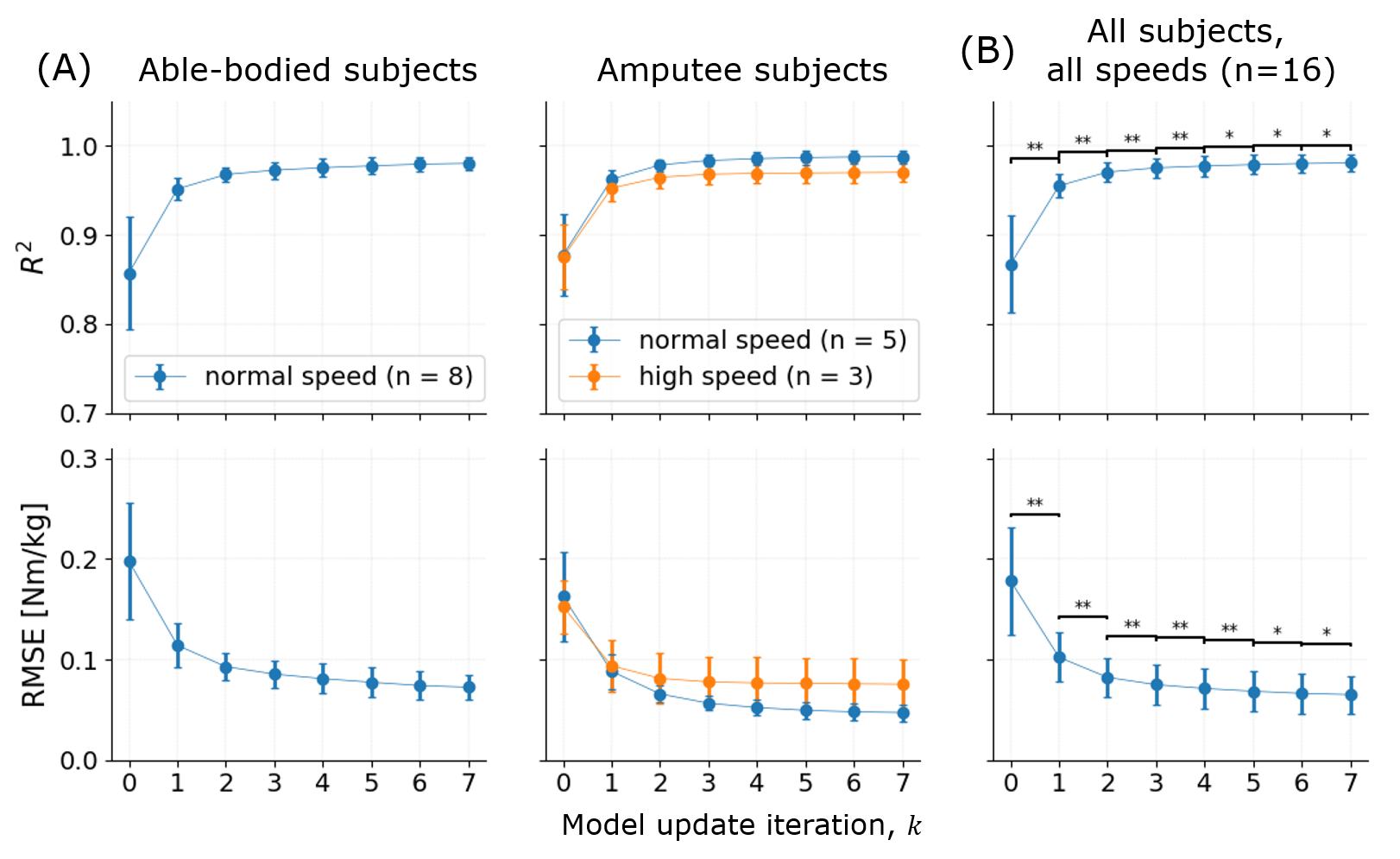}
    \caption{(A) Left: Mean $R^2$ and \textit{RMSE} for $\tau_{ankle}$ predictions by hybrid models for normal speed walking datasets of eight able bodied subjects. Right: Mean $R^2$ and \textit{RMSE} for $\tau_{ankle}$ predictions by hybrid models for normal speed walking datasets of five transtibial amputee subjects and higher speed walking datasets of three of the five transtibial amputees. The horizontal axis indicates the index, $k$, of model update iteration ($k=0$ indicates baseline inter-individual model, $f^{II}$ or $f^{hybrid}_0$, and $k=7$ indicates $f^{hybrid}_7$. For amputees, the predictions on high-speed walking trials (orange curves on the right pane) were obtained by refining the model using only the data from normal speed walking trials. The error bars are one standard deviation from the mean). (B) $R^2$ and \textit{RMSE} for $\tau_{ankle}$ predictions averaged across normal speed walking trials of able-bodied and normal and high speed walking trials of amputee subjects (n=16) during each hybrid model update iteration. The results of Wilcoxon-signed-rank test comparing the performance of successive update iterations is also shown ($^* p<1e-2, ^{**} p<1e-3$). }
    \label{fig:hybrid_amputee_r2_rmse}
\end{figure*}

Fig. \ref{fig:hybrid_amputee_r2_rmse} (A) shows the mean $R^2$ and \textit{RMSE} for the  mass normalized $\tau_{ankle}$ out-of-sample predictions by the hybrid model on normal speed trials for eight able-bodied subjects for normal speed walking, normal speed walking trials of five amputee subjects, and high speed walking trials of Amputees 1, 2, and 3. The predictions on the high-speed walking trials for the amputees were from the hybrid model trained with only normal speed walking trials, i.e., the high-speed walking trials were never a part of the training set. 
 
In general, the accuracy of  $\tau_{ankle}$ predictions of able-bodied and amputee subjects increased monotonically with the addition of individual-specific data for training although the accuracy plateaued after the addition of three to four gait cycles of training data. For the able-bodied subjects, the baseline model gave a mean $R^2$ of 0.86 (mean RMSE = 0.2Nm/kg). After seven iterations of individual-specific training (selective update with a threshold of 0.99), the mean $R^2$ score reached 0.98 (mean RMSE = 0.07Nm/kg). For the amputee subjects, the baseline inter-individual model gave comparable accuracy for both normal and high speed walking trials (mean $R^2 ~ 0.88$ and mean RMSE $<$ 0.16Nm/kg). The accuracy of  $\tau_{ankle}$ predictions increased with the addition of individual-specific normal speed training data. After seven iterations of individual-specific update (selective update with 0.99 threshold) an $R^2$ of 0.99 and RMSE of 0.05Nm/kg was obtained for torque predictions on out-of-sample normal walking trials. Corresponding mean $R^2$ and mean RMSE for high-speed walking trials were 0.97 and 0.08 respectively. 

\begin{figure}
    \centering
    \includegraphics[width =  0.4\textwidth]{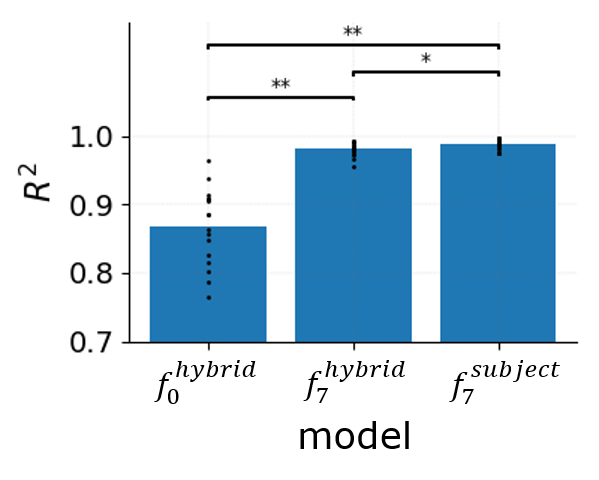}
    \caption{Comparison of $R^2$ values of the baseline inter-individual model, a hybrid model after the seventh individual-specific iteration, and a pure individual specific model trained using seven individual-specific trials. Statistical comparisons between the models were performed using Wilcoxon-signed-rank test ($^* p<1e-2, ^{**} p<1e-3$).}
    \label{fig:stat_comparison_models}
\end{figure}

A paired statistical comparison across normal speed trials for all subjects (eight able-bodied and five transtibial amputees) showed that the performance of the hybrid model improved significantly during each iteration of the individual-specific training (Fig. \ref{fig:hybrid_amputee_r2_rmse} (B)). Consequently, the hybrid model after the seventh individual-specific iteration, $f_7^{hybrid}$ performed significantly better than the baseline inter-individual model, $f_0^{hybrid}$ (Fig. \ref{fig:stat_comparison_models}). Even though a pure individual-specific model, $f_7^{subject}$, performed slightly better than the hybrid model, the percentage improvement was only 1\%. The slight difference in accuracy could be attributed to the effect of baseline inter-individual training of the hybrid model. We assume that the difference will further narrow if more individual-specific trials are used for updating the hybrid model. 

\begin{figure*}[!htbp]
    \centering
    \includegraphics[width = 0.98\textwidth]{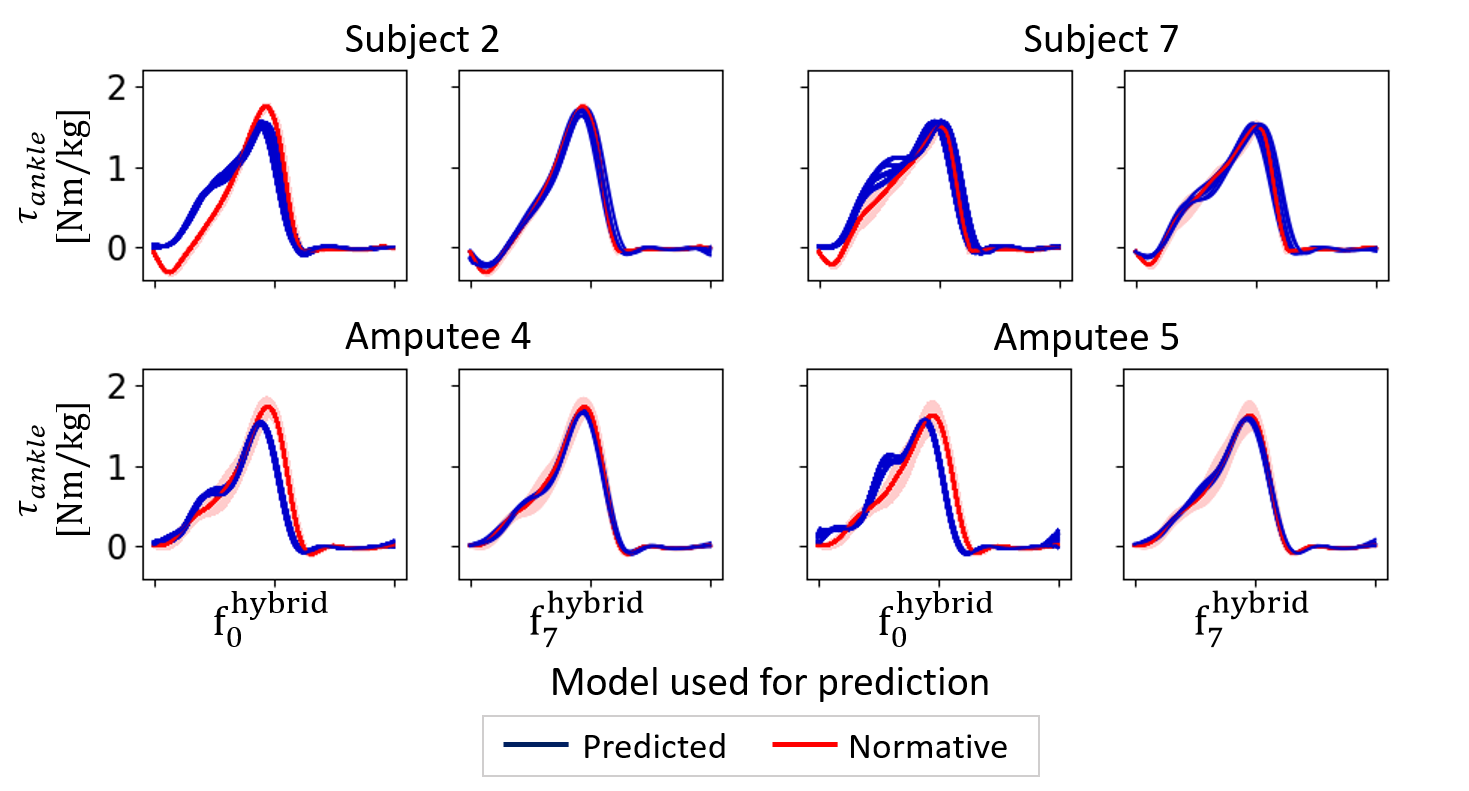}
    \caption{$\tau_{ankle}$ trajectories predicted by the hybrid model for normal speed walking trials of able-bodied subjects 2 and 7, and Amputees 4 and 5, using baseline inter-individual model, $f_0^{hybrid}$ and hybrid model, $f_7^{hybrid}$, after seventh individual-specific training iteration. The multiple blue curves are the predicted $\tau_{ankle}$ trajectories for all the cross-validation iterations. The red curves represent the median measured trajectories (read normative trajectory for amputees)  with one median absolute deviation (shaded area).}
    \label{fig:hybrid_amp_pred}
\end{figure*}

Fig. \ref{fig:hybrid_amp_pred} shows the predicted trajectories of $\tau_{ankle}$ for normal and fast walking speed for two able-bodied subjects (Subject 2 and 7) and two transtibial amputee subjects (Amputees 4 and 5) using the baseline inter-individual model ($f_0^{hybrid}$) and the hybrid model trained using individual-specific normal speed gait data after the seventh update iteration ($f_7^{hybrid}$). Initial predictions by the baseline inter-individual model appeared slightly impaired and highly variable for different subjects. With individual-specific normal walk training, the deviations between predictions and target normative trajectories appeared to improve (e.g., better alignment of peak plantarflexor moment). 

\begin{figure*}[!hb]
    \centering
    \includegraphics[width=\textwidth]{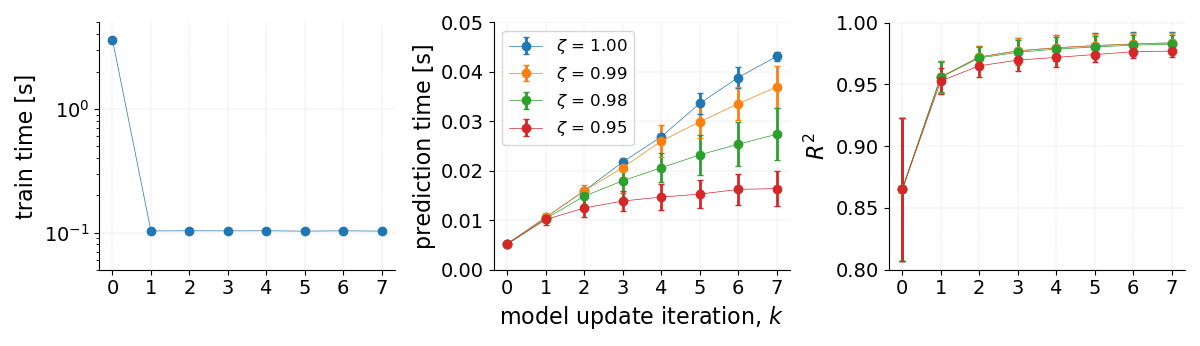}
    \caption{Training times, prediction times, and $R^2$ for each iteration of the hybrid model update for different threshold values ($\zeta$) used for selective update of the model. Each datapoint was obtained by averaging across normal speed trials of all subjects (n=13). The horizontal axis indicates the index of individual-specific update iteration, $k$ ($k=0$ indicate baseline inter-individual model). The error bars indicate one standard deviation from the mean. }
    \label{fig:time_complexity}
\end{figure*}

Fig. \ref{fig:time_complexity} shows the time required for training and prediction and the $R^2$ score of the hybrid model during each iteration for different threshold values for a selective update. The highest training time was required for the baseline inter-individual model as it was trained using the largest amount of data (data from five trials each of 30 subjects). The time required for each individual-specific training iteration remained consistent at around 100ms (learning rate of 2kHz). On the other hand, the prediction time for each trial (containing 200 samples) increased with each iteration of the individual-specific training. A threshold value $\zeta=1$ led to a linear increase in prediction time with each update iteration. By lowering the threshold, the prediction times decreased with an increasing trend towards saturation (Fig. \ref{fig:time_complexity}). The prediction accuracy decreased only negligibly by lowering the threshold. For a $\zeta$ value of 0.95, the mean $R^2$ across 13 subjects stayed at 0.98 (0.7\% decrease compared to $\zeta=1$), while the prediction time per trial was 16ms (compared to 43ms with $\zeta$=1) after seven individual-specific iterations. This corresponds to a prediction rate larger than 10kHz.
The learning and prediction rates are well within the range of real-time requirements \cite{nguyen2009local}. 

\section{Discussion} \label{sec:discussion}
 
In this study, the plausibility of a dynamic torque prediction model for below-knee amputees was explored by incremental inclusion of estimators to predict the normative ankle joint moment values from the subject’s residual limb kinematics. Tree-based ensembles was chosen as the torque prediction model for the possibility of dynamically adjusting the number of estimators without retraining the complete model. 

A hybrid of two approaches (a generic inter-individual and individual-specific adaptation model) was used. 
Even though a generically trained inter-individual model could be directly applied for torque prediction on new subjects, it cannot cater to the individual specific variations in gait patterns. On the other hand, training an individual-specific model requires acquiring enough training data from each new subject, tuning hyper-parameters of the model, and training and validating the model before it could be applied for torque prediction. By combining the two approaches, the hybrid model attempts to provide the advantages of both inter-individual and individual-specific models, while ameliorating their limitations. 

The baseline inter-individual model was trained using a public dataset containing trials from five speed levels of 30 able-bodied subjects. Our choice of such a heterogeneous dataset was motivated by the aim to develop a generalized model that can accommodate several speeds that the amputee may choose to walk at. This was reflected to some extent from the comparable accuracy of prediction for both normal and fast walking speeds for Amputees 2 and 3 while using the baseline inter-individual model (Fig. \ref{fig:hybrid_amputee_r2_rmse}). However, the torque predictions using the baseline inter-individual model were relatively inaccurate and highly varied across different subjects (Fig. \ref{fig:hybrid_amputee_r2_rmse} and \ref{fig:hybrid_amp_pred}). 

Further, the hybrid model takes advantage of the individual-specific approach to improve its prediction performance for the particular subject by dynamically updating the model using individual-specific data. With the addition of new estimators fitted to individual-specific data, the accuracy quickly increased and almost plateaued after two to three iterations of individual-specific training (Fig. \ref{fig:hybrid_amputee_r2_rmse}), reaching an accuracy level similar to a purely individual-specific model (Fig. \ref{fig:stat_comparison_models},  \cite{dey2020feasibility}). Correspondingly, the discrepancy between the target and predicted trajectories of $\tau_{ankle}$ was normalized (Fig. \ref{fig:hybrid_amp_pred}). This sustains from our previous study, where we showed that individual-specific gait data were more accurate in learning a relation between corresponding input features and target variables than an inter-individual model \cite{dey2020feasibility}. The convergence of the hybrid model for amputees shows that as the hybrid model was updated by more amputee-specific patterns, it could predict normative torque patterns more accurately, even though the subject’s input gait patterns were affected by amputations, which is in confirmation with our initial hypothesis. 
For the amputee subjects, the proposed approach also solves the problem of incomplete training data for active prosthesis control by learning a mapping from the amputee specific input trajectories obtained from the residual limb to a normative ankle torque trajectory calculated from a sample of able-bodied individuals. 

The accuracy of predictions obtained using our method after individual-specific updates is comparable to or better than that of the previous studies which used pre-trained regression algorithms for gait variable predictions \cite{ardestani2014human,dey2019support,dey2020feasibility,xiong2019intelligent,mundt2020estimation} (see table \ref{tab:comparison}). Even though we have used the ankle torque as the prediction variable in this study, the training approach that we proposed may be applicable to other high-level variables probably either having similar complexity in the time domain or are functions of joint moments or angles. This means that our algorithm will likely achieve similar accuracy even if it had to predict a different variable (e.g., joint velocity).

\begin{table*}[!htbp]
\centering
\caption{Comparison of performance of our models with those used in other relevant studies}
\label{tab:comparison}
\begin{tabular}{cccccc} 
\toprule
Study & Model used & Input features & Predicted variable & Prediction accuracy & Our accuracy \\ 
\midrule
 \cite{ardestani2014human}& Wavelet neural network & GRF, EMG & Joint moments & NRMSE~\textless{} 10\%,~$\rho$ = 0.94 & \multirow{6}{*}{\begin{tabular}[c]{@{}c@{}}$R^2$ = 0.98,~\\RMSE $<$ 0.07Nm/kg\end{tabular}} \\
 \cite{xiong2019intelligent}& Artificial neural network & EMG, joint angles & Joint moments & NRMSE \textless{}
 7.89\%, $\rho$~= 0.9633 &  \\
 \cite{mundt2020estimation}& Artificial neural network & IMU data & \begin{tabular}[c]{@{}c@{}}Joint angles, \\moments\end{tabular} & \begin{tabular}[c]{@{}c@{}}Joint angles:~RMSE \textless{} 4.8º,~$\rho$~= 0.85\\Joint moments:~nRMSE \textless{} 13\%, $\rho$~= 0.95\end{tabular} &  \\
 \cite{dey2019support}& Support vector regression & \begin{tabular}[c]{@{}c@{}}Joint angles, \\segment orientations\end{tabular} & \begin{tabular}[c]{@{}c@{}}Ankle angle, \\ankle moment\end{tabular} & \begin{tabular}[c]{@{}c@{}}~$R^2$~= 0.97 ($\tau_{ankle}$)\\~$R^2$~= 0.98 ($\theta_{ankle}$)\end{tabular} &  \\
 \cite{yun2014statistical}& Gaussian process regression & \begin{tabular}[c]{@{}c@{}}Gait cycle time, \\anthropometry\end{tabular} & Joint kinematics & Mean error \textless{} 4.3º ($\theta_{ankle}$) &  \\
 \cite{dey2020feasibility}& Random Forest regression & \begin{tabular}[c]{@{}c@{}}Joint angles, \\segment orientations\end{tabular} & \begin{tabular}[c]{@{}c@{}}Ankle angle, \\ankle moment\end{tabular} & \begin{tabular}[c]{@{}c@{}}~$R^2$ = 0.97 ($\tau_{ankle}$)\textit{~}\\~$R^2$ = 0.98 ($\theta_{ankle}$)\end{tabular} &  \\
\bottomrule
\end{tabular}
\end{table*}

For usage in a real application (Fig. \ref{fig:schematic}), the control model for the active prosthesis could be pre-trained using data from able-bodied subjects. The inputs from the prosthesis user should be acquired from wearable sensors like inertial measurement units and goniometers. The signals should be processed (eg. filtering) in real-time and the relevant features should be extracted. The model predicts the output ($\tau_{ankle}$) from the input features and a low-level controller takes as input, the predicted $\tau_{ankle}$ values to execute torque control of the prosthetic ankle joint. Once the user starts ambulating using the powered prosthesis controlled using the pre-trained model, new training data could be collected and the model could be incrementally updated. We assume that this update may gradually compensate for the altered input patterns due to the user adapting to the prosthesis during ambulation.
On the other hand, training a pure individual-specific model would require the new subject to walk with a passive prosthesis during training data acquisition, . As a result, the trained model would not be able to take into account the changes in the interaction between the user and the prosthesis when the user walks with a powered prosthesis.

A practical concern of such iterative training is that it would keep on increasing the number of estimators in the model, in turn increasing the prediction delays. To prevent an explosion of the number of estimators, a selective update and drop policy could be defined. A selective update policy would be to update the model only when the predicted trajectories vary substantially from the normative trajectory. In this study, we used a selective update policy where new estimators are fitted if the prediction accuracy ($R^2$) is below a threshold of 0.99. We also observed that by lowering the threshold value it was possible to reduce the prediction delays with only a negligible decrease in accuracy (Fig. \ref{fig:time_complexity}). Additionally, a drop policy could also be used to drop correlated estimators. In some cases, it may also be helpful to keep only those estimators that were fitted for recent locomotion data and others discarded. Furthermore, more heuristics could be applied to constrain updating the model with data that are within a pre-defined range of normative inputs to avoid model training with extremely noisy data.

In this study, we have established through an offline analysis, the possibility of the proposed hybrid approach to adapt to new training inputs and thereby adjust its prediction to converge towards normative values with fairly few training samples. An online analysis of the stability of gait remains a crucial part of our future studies. Specifically, since the prediction models were trained using pre-recorded data, it remains to be analyzed how the predicted gait variables affect the upcoming input patterns and how the model would handle such changes. For example, at the beginning of the training when the predictions come from the baseline model, it is likely that the user show adaptations in his gait due to the yet not optimized ankle torque. It needs to be investigated how such adaptations when used for training the algorithm  would affect subsequent gait cycles. It would also be interesting to examine how the proposed hybrid approach adapts to other common locomotion modes such as stair ambulation. As the amount of individual-specific training data and locomotion modes increase, effective policies for retraining and discarding decision estimators will presumably become necessary to optimize the efficiency of the model while maintaining its flexibility. Additionally, we performed all our analyses in this study using a powerful laptop PC. The feasibility of such a dynamic iterative training on low power micro-controllers should also be investigated. Furthermore, future studies should apply the hybrid model on a much larger sample of transtibial amputees.

\section{Conclusion} \label{sec:conclusion}
We have proposed here a hybrid strategy for dynamically training a torque estimation model from biologically inspired temporal patterns of the ankle joint torque during gait. For this, we created a hybrid of two different models (inter-individual and individual-specific) by an iterative training procedure to predict normative ankle joint moment from the hip, knee, and shank kinematics during level-ground walking for able-bodied subjects and unilateral below-knee amputees. It was found that the model swiftly converges to high prediction accuracy as estimators are fitted to individual-specific data. The proposed approach also solves the challenge of training a prediction model for amputees using incomplete user-specific training data. Furthermore, our approach may be applied to predict other high-level variables probably either having similar complexity in the time domain or are functions of joint moments or angles. Our findings indicate the potential of the proposed hybrid approach for developing a robust torque control strategy for active ankle prostheses/orthoses.

\section*{Acknowledgment}

We immensely thank Dr. Marko Markovic, Applied   Rehabilitation  Technology  Lab   (ART-Lab), University Medical Center Goettingen,   Goettingen, Germany for the valuable feedback that enhanced the quality of the manuscript. We also thank Ms. Antonia Bernecker, Applied   Rehabilitation  Technology  Lab   (ART-Lab), University Medical Center Goettingen, Goettingen, Germany, for partly assisting with the data organization. This study was partly supported by the grant (INOPRO-16SV7656) from the German Federal Ministry of Education and Research.

\ifCLASSOPTIONcaptionsoff
  \newpage
\fi

\bibliographystyle{ieeetr}
\bibliography{reference}

\begin{thebibliography}{10}

\bibitem{au2008powered}
S.~Au, M.~Berniker, and H.~Herr, ``Powered ankle-foot prosthesis to assist
  level-ground and stair-descent gaits,'' {\em Neural Networks}, vol.~21,
  no.~4, pp.~654--666, 2008.

\bibitem{donelan2002mechanical}
J.~M. Donelan, R.~Kram, and A.~D. Kuo, ``Mechanical work for step-to-step
  transitions is a major determinant of the metabolic cost of human walking,''
  {\em Journal of Experimental Biology}, vol.~205, no.~23, pp.~3717--3727,
  2002.

\bibitem{neptune2001contributions}
R.~R. Neptune, S.~Kautz, and F.~Zajac, ``Contributions of the individual ankle
  plantar flexors to support, forward progression and swing initiation during
  walking,'' {\em Journal of biomechanics}, vol.~34, no.~11, pp.~1387--1398,
  2001.

\bibitem{winter1983energy}
D.~A. Winter, ``Energy generation and absorption at the ankle and knee during
  fast, natural, and slow cadences,'' {\em Clinical Orthopaedics and Related
  Research{\textregistered}}, vol.~175, pp.~147--154, 1983.

\bibitem{waters1976energy}
R.~Waters, J.~Perry, D.~Antonelli, and H.~Hislop, ``Energy cost of walking of
  amputees: the influence of level of amputation,'' {\em J Bone Joint Surg Am},
  vol.~58, no.~1, pp.~42--46, 1976.

\bibitem{seroussi1996mechanical}
R.~E. Seroussi, A.~Gitter, J.~M. Czerniecki, and K.~Weaver, ``Mechanical work
  adaptations of above-knee amputee ambulation,'' {\em Archives of physical
  medicine and rehabilitation}, vol.~77, no.~11, pp.~1209--1214, 1996.

\bibitem{herr2012bionic}
H.~M. Herr and A.~M. Grabowski, ``Bionic ankle--foot prosthesis normalizes
  walking gait for persons with leg amputation,'' {\em Proceedings of the Royal
  Society B: Biological Sciences}, vol.~279, no.~1728, pp.~457--464, 2012.

\bibitem{au2009powered}
S.~K. Au, J.~Weber, and H.~Herr, ``Powered ankle--foot prosthesis improves
  walking metabolic economy,'' {\em IEEE Transactions on Robotics}, vol.~25,
  no.~1, pp.~51--66, 2009.

\bibitem{cherelle2013design}
P.~Cherelle, V.~Grosu, A.~Matthys, B.~Vanderborght, and D.~Lefeber, ``Design
  and validation of the ankle mimicking prosthetic (amp-) foot 2.0,'' {\em IEEE
  Transactions on Neural Systems and Rehabilitation Engineering}, vol.~22,
  no.~1, pp.~138--148, 2013.

\bibitem{hitt2010active}
J.~K. Hitt, T.~G. Sugar, M.~Holgate, and R.~Bellman, ``An active foot-ankle
  prosthesis with biomechanical energy regeneration,'' {\em Journal of medical
  devices}, vol.~4, no.~1, 2010.

\bibitem{zhuang2019ensemble}
J.~Zhuang, K.~Geng, and G.~Yin, ``Ensemble learning based brain-computer
  interface system for ground vehicle control,'' {\em IEEE Transactions on
  Systems, Man, and Cybernetics: Systems}, 2019.

\bibitem{hahne2018simultaneous}
J.~M. Hahne, M.~A. Schweisfurth, M.~Koppe, and D.~Farina, ``Simultaneous
  control of multiple functions of bionic hand prostheses: Performance and
  robustness in end users,'' {\em Science Robotics}, vol.~3, no.~19,
  p.~eaat3630, 2018.

\bibitem{varol2007decomposition}
H.~A. Varol and M.~Goldfarb, ``Decomposition-based control for a powered knee
  and ankle transfemoral prosthesis,'' in {\em 2007 IEEE 10th International
  Conference on Rehabilitation Robotics}, pp.~783--789, IEEE, 2007.

\bibitem{sup2009preliminary}
F.~Sup, H.~A. Varol, J.~Mitchell, T.~J. Withrow, and M.~Goldfarb, ``Preliminary
  evaluations of a self-contained anthropomorphic transfemoral prosthesis,''
  {\em IEEE/ASME Transactions on mechatronics}, vol.~14, no.~6, pp.~667--676,
  2009.

\bibitem{sup2010upslope}
F.~Sup, H.~A. Varol, and M.~Goldfarb, ``Upslope walking with a powered knee and
  ankle prosthesis: initial results with an amputee subject,'' {\em IEEE
  transactions on neural systems and rehabilitation engineering}, vol.~19,
  no.~1, pp.~71--78, 2010.

\bibitem{huang2011continuous}
H.~Huang, F.~Zhang, L.~J. Hargrove, Z.~Dou, D.~R. Rogers, and K.~B. Englehart,
  ``Continuous locomotion-mode identification for prosthetic legs based on
  neuromuscular--mechanical fusion,'' {\em IEEE Transactions on Biomedical
  Engineering}, vol.~58, no.~10, pp.~2867--2875, 2011.

\bibitem{varol2009multiclass}
H.~A. Varol, F.~Sup, and M.~Goldfarb, ``Multiclass real-time intent recognition
  of a powered lower limb prosthesis,'' {\em IEEE Transactions on Biomedical
  Engineering}, vol.~57, no.~3, pp.~542--551, 2009.

\bibitem{young2014intent}
A.~J. Young, A.~M. Simon, N.~P. Fey, and L.~J. Hargrove, ``Intent recognition
  in a powered lower limb prosthesis using time history information,'' {\em
  Annals of biomedical engineering}, vol.~42, no.~3, pp.~631--641, 2014.

\bibitem{au2005emg}
S.~K. Au, P.~Bonato, and H.~Herr, ``An emg-position controlled system for an
  active ankle-foot prosthesis: an initial experimental study,'' in {\em
  Rehabilitation robotics, 2005. ICORR 2005. 9th international conference on},
  pp.~375--379, IEEE, 2005.

\bibitem{huang2008strategy}
H.~Huang, T.~A. Kuiken, R.~D. Lipschutz, {\em et~al.}, ``A strategy for
  identifying locomotion modes using surface electromyography,'' {\em IEEE
  Transactions on Biomedical Engineering}, vol.~56, no.~1, pp.~65--73, 2008.

\bibitem{8302866}
D.~{Quintero}, D.~J. {Villarreal}, D.~J. {Lambert}, S.~{Kapp}, and R.~D.
  {Gregg}, ``Continuous-phase control of a powered knee–ankle prosthesis:
  Amputee experiments across speeds and inclines,'' {\em IEEE Transactions on
  Robotics}, vol.~34, no.~3, pp.~686--701, 2018.

\bibitem{villarreal2020controlling}
D.~J. Villarreal and R.~D. Gregg, ``Controlling a powered transfemoral
  prosthetic leg using a unified phase variable,'' in {\em Wearable Robotics},
  pp.~487--506, Elsevier, 2020.

\bibitem{embry2018modeling}
K.~R. Embry, D.~J. Villarreal, R.~L. Macaluso, and R.~D. Gregg, ``Modeling the
  kinematics of human locomotion over continuously varying speeds and
  inclines,'' {\em IEEE transactions on neural systems and rehabilitation
  engineering}, vol.~26, no.~12, pp.~2342--2350, 2018.

\bibitem{horst2019explaining}
F.~Horst, S.~Lapuschkin, W.~Samek, K.-R. M{\"u}ller, and W.~I. Sch{\"o}llhorn,
  ``Explaining the unique nature of individual gait patterns with deep
  learning,'' {\em Scientific reports}, vol.~9, no.~1, p.~2391, 2019.

\bibitem{ardestani2014human}
M.~M. Ardestani, X.~Zhang, L.~Wang, Q.~Lian, Y.~Liu, J.~He, D.~Li, and Z.~Jin,
  ``Human lower extremity joint moment prediction: A wavelet neural network
  approach,'' {\em Expert Systems with Applications}, vol.~41, no.~9,
  pp.~4422--4433, 2014.

\bibitem{xiong2019intelligent}
B.~Xiong, N.~Zeng, H.~Li, Y.~Yang, Y.~Li, M.~Huang, W.~Shi, M.~Du, and
  Y.~Zhang, ``Intelligent prediction of human lower extremity joint moment: an
  artificial neural network approach,'' {\em IEEE Access}, vol.~7,
  pp.~29973--29980, 2019.

\bibitem{lim2020prediction}
H.~Lim, B.~Kim, and S.~Park, ``Prediction of lower limb kinetics and kinematics
  during walking by a single imu on the lower back using machine learning,''
  {\em Sensors}, vol.~20, no.~1, p.~130, 2020.

\bibitem{mundt2020estimation}
M.~Mundt, A.~Koeppe, S.~David, T.~Witter, F.~Bamer, W.~Potthast, and
  B.~Markert, ``Estimation of gait mechanics based on simulated and measured
  imu data using an artificial neural network,'' {\em Frontiers in
  Bioengineering and Biotechnology}, vol.~8, 2020.

\bibitem{dey2019support}
S.~Dey, M.~Eslamy, T.~Yoshida, M.~Ernst, T.~Schmalz, and A.~Schilling, ``A
  support vector regression approach for continuous prediction of ankle angle
  and moment during walking: An implication for developing a control strategy
  for active ankle prostheses,'' in {\em 2019 IEEE 16th International
  Conference on Rehabilitation Robotics (ICORR)}, pp.~727--733, IEEE, 2019.

\bibitem{dhir2018locomotion}
N.~Dhir, H.~Dallali, E.~M. Ficanha, G.~A. Ribeiro, and M.~Rastgaar,
  ``Locomotion envelopes for adaptive control of powered ankle prostheses,'' in
  {\em 2018 IEEE International Conference on Robotics and Automation (ICRA)},
  pp.~1488--1495, IEEE, 2018.

\bibitem{yun2014statistical}
Y.~Yun, H.-C. Kim, S.~Y. Shin, J.~Lee, A.~D. Deshpande, and C.~Kim,
  ``Statistical method for prediction of gait kinematics with gaussian process
  regression,'' {\em Journal of Biomechanics}, vol.~47, no.~1, pp.~186--192,
  2014.

\bibitem{allard2017urban}
P.~Allard, S.~Leteneur, {\'E}.~Watelain, and M.~Begon, ``Urban legends in gait
  analysis,'' {\em Movement \& Sport Sciences-Science \& Motricit{\'e}},
  no.~98, pp.~5--11, 2017.

\bibitem{wahid2016multiple}
F.~Wahid, R.~Begg, N.~Lythgo, C.~J. Hass, S.~Halgamuge, and D.~C. Ackland, ``A
  multiple regression approach to normalization of spatiotemporal gait
  features,'' {\em Journal of applied biomechanics}, vol.~32, no.~2,
  pp.~128--139, 2016.

\bibitem{stansfield2003normalisation}
B.~Stansfield, S.~Hillman, M.~Hazlewood, A.~Lawson, A.~Mann, I.~Loudon, and
  J.~Robb, ``Normalisation of gait data in children,'' {\em Gait \& posture},
  vol.~17, no.~1, pp.~81--87, 2003.

\bibitem{hof1996scaling}
A.~L. Hof, ``Scaling gait data to body size,'' {\em Gait \& posture}, vol.~3,
  no.~4, pp.~222--223, 1996.

\bibitem{senden2012importance}
R.~Senden, K.~Meijer, I.~Heyligers, H.~Savelberg, and B.~Grimm, ``Importance of
  correcting for individual differences in the clinical diagnosis of gait
  disorders,'' {\em Physiotherapy}, vol.~98, no.~4, pp.~320--324, 2012.

\bibitem{breiman2001random}
L.~Breiman, ``Random forests,'' {\em Machine learning}, vol.~45, no.~1,
  pp.~5--32, 2001.

\bibitem{quinlan1986induction}
J.~R. Quinlan, ``Induction of decision trees,'' {\em Machine learning}, vol.~1,
  no.~1, pp.~81--106, 1986.

\bibitem{Schaal2006}
S.~Schaal, {\em Dynamic Movement Primitives -A Framework for Motor Control in
  Humans and Humanoid Robotics}, pp.~261--280.
\newblock Tokyo: Springer Tokyo, 2006.

\bibitem{mukovskiy2017}
A.~Mukovskiy, C.~Vassallo, M.~Naveau, O.~Stasse, P.~Souères, and M.~Giese,
  ``Adaptive synthesis of dynamically feasible full-body movements for the
  humanoid robot hrp-2 by flexible combination of learned dynamic movement
  primitives,'' {\em Robotics and Autonomous Systems}, vol.~online, 02 2017.

\bibitem{carnier2021ml}
R.~Matos~Carnier and Y.~Fujimoto, ``Assessment of machine learning of optimal
  solutions for robotic walking,'' {\em International Journal of Mechanical
  Engineering and Robotics Research}, vol.~10, pp.~44--48, 01 2021.

\bibitem{fukuchi2018public}
C.~A. Fukuchi, R.~K. Fukuchi, and M.~Duarte, ``A public dataset of overground
  and treadmill walking kinematics and kinetics in healthy individuals,'' {\em
  PeerJ}, vol.~6, p.~e4640, 2018.

\bibitem{Opensim:Delp:open-source}
S.~L. Delp, F.~C. Anderson, A.~S. Arnold, P.~Loan, A.~Habib, T.~John,
  E.~Guendelman, and D.~G. Thelen, ``Opensim: Open-source software to create
  and analyze dynamic simulations of movement.''

\bibitem{riley2007kinematic}
P.~O. Riley, G.~Paolini, U.~Della~Croce, K.~W. Paylo, and D.~C. Kerrigan, ``A
  kinematic and kinetic comparison of overground and treadmill walking in
  healthy subjects,'' {\em Gait \& posture}, vol.~26, no.~1, pp.~17--24, 2007.

\bibitem{breiman1984classification}
L.~Breiman, J.~Friedman, R.~Olshen, and C.~Stone, ``Classification and
  regression trees. wadsworth int,'' {\em Group}, vol.~37, no.~15,
  pp.~237--251, 1984.

\bibitem{nguyen2009local}
D.~Nguyen-Tuong, J.~R. Peters, and M.~Seeger, ``Local gaussian process
  regression for real time online model learning,'' in {\em Advances in Neural
  Information Processing Systems}, pp.~1193--1200, 2009.

\bibitem{dey2020feasibility}
S.~Dey, T.~Yoshida, and A.~Schilling, ``Feasibility of training a random forest
  model with incomplete user-specific data for devising a control strategy for
  active biomimetic ankle,'' {\em Frontiers in Bioengineering and
  Biotechnology}, vol.~8, p.~855, 2020.

\end{thebibliography}

\end{document}